\documentclass[pdflatex,sn-basic]{sn-jnl}

\usepackage{graphicx}%
\usepackage{multirow}%
\usepackage{amsmath,amssymb,amsfonts}%
\usepackage{amsthm}%
\usepackage{mathrsfs}%
\usepackage[title]{appendix}%
\usepackage{xcolor}%
\usepackage{textcomp}%
\usepackage{manyfoot}%
\usepackage{booktabs}%
\usepackage{algorithm}%
\usepackage{algorithmicx}%
\usepackage{algpseudocode}%
\usepackage{listings}%

\theoremstyle{thmstyleone}%
\theoremstyle{thmstyletwo}%

\theoremstyle{thmstylethree}%

\usepackage[T1]{fontenc}    
\usepackage{hyperref}       
\usepackage{url}            
\usepackage{graphicx}
\usepackage{siunitx}
\usepackage{xcolor}
\usepackage{comment}

\usepackage{graphicx}
\usepackage{amsmath}
\usepackage{pifont}
\usepackage{amsmath, amssymb, amsfonts, amsthm}
\usepackage{graphicx}
\usepackage{amsfonts}       
\usepackage{nicefrac}       
\usepackage{microtype}      
\usepackage{mathtools}
\usepackage{comment}
\usepackage{pbox}
\usepackage{textcomp}
\usepackage[utf8]{inputenc}
\usepackage{multirow}
\usepackage{nameref}
\usepackage{textalpha}
\usepackage{booktabs} 
\usepackage{comment}
\usepackage{subfig}
\usepackage{threeparttable}
\usepackage{caption}
\usepackage{mathrsfs}
\usepackage{tabularx}

\usepackage{ragged2e} 
\usepackage{makecell}
\usepackage{colortbl}
\usepackage{float}
\usepackage{placeins}
\usepackage{flafter}
\usepackage{rotating}
\usepackage[explicit]{titlesec}
\usepackage{longtable}
\usepackage{cleveref}

\usepackage[normalem]{ulem} 

\begin{document}

\title{Towards actionable hypotension prediction-- predicting catecholamine therapy initiation in the intensive care unit}

\author[1]{\fnm{Richard} \sur{Koebe}}\email{koebe.richard@klinikum-oldenburg.de}
\author[2]{\fnm{Noah} \sur{Saibel}}\email{noah.saibel@uol.de}
\author[2]{\fnm{Juan Miguel} \sur{Lopez Alcaraz}}\email{juan.lopez.alcaraz@uol.de}
\author[1]{\fnm{Simon} \sur{Schäfer}}\email{schaefer.simon@klinikum-oldenburg.de}
\author[2]{\fnm{Nils} \sur{Strodthoff}}\email{nils.strodthoff@uol.de}

\equalcont{Corresponding author. These authors contributed equally (first authors): Richard Koebe and Noah Saibel. These authors contributed equally (senior authors): Simon Schäfer and Nils Strodthoff.}

\affil[1]{\orgdiv{University Clinic of Anesthesiology, Intensive Care Medicine, Emergency Medicine, and Pain Therapy}, \orgname{Klinikum Oldenburg}, \orgaddress{\street{Rahel-Straus-Straße 10}, \city{Oldenburg}, \postcode{26133}, \state{Lower-Saxony}, \country{Germany}}}

\affil[2]{\orgdiv{AI4Health Division}, \orgname{Carl von Ossietzky Universität}, \orgaddress{\street{Ammerländer Heerstr. 114-118}, \city{Oldenburg}, \postcode{26129}, \state{Lower-Saxony}, \country{Germany}}}


\abstract{

\textbf{Background:} In critically ill patients hypotension is common and potentially life-threatening. Initiation of catecholamine therapy marks a critical management step, with both undertreatment and overtreatment posing risks. Hypotension in intensive care unit (ICU) patients is common and potentially life-threatening. Escalation to catecholamine therapy marks a critical management step, with both undertreatment and overtreatment posing risks. Most machine learning methods predict hypotension via fixed MAP thresholds or MAP forecasting, but these overlook the clinical decision-making behind treatment escalation. Predicting catecholamine initiation defined here as the start of administration of vasoactive or inotropic agents, thus offers a clinically actionable target that better reflects this decision.

\textbf{Methods:} We used the Medical Information Mart for Intensive Care III (MIMIC-III) database to investigate catecholamine initiation in the ICU. The binary target was defined as initiation within a 15-minute prediction window. As input features, we extracted statistical measures from a sliding two-hour MAP context window preceding the prediction window. In addition, we incorporated demographics, biometrics, comorbidities, and ongoing treatments. We trained an Extreme Gradient Boosting (XGBoost) model and applied SHapley Additive exPlanations (SHAP) to interpret feature importance.
 
\textbf{Results:} The model achieved strong discrimination with an area under the receiver operating curve (AUROC) of 0.822 (0.813-0.830), outperforming a baseline defined by hypotension (MAP < 65), which yielded 0.686 (0.675-0.699). SHAP feature importance analysis identified recent MAP values, MAP trends, and ongoing treatments such as sedatives and electrolytes as the dominant predictors. Subgroup analyses showed above-average performance in males, younger patients (age < 53), those with higher BMI (> 32), and usually patients without comorbidities nor other current drugs administered.

\textbf{Conclusion:} Predicting catecholamine initiation from MAP dynamics, ongoing therapy and patient context addresses the critical question of when to escalate treatment. This shifts the focus from threshold-based hypotension alarms toward supporting the complex decision of when treatment escalation is warranted. Our results demonstrate the feasibility of this approach for a broad ICU cohort under the natural imbalance of initiation events. Future work should incorporate richer temporal and physiological context, extend the label definition beyond initiation onsets to also capture the escalation of ongoing catecholamine therapies and benchmark performance against existing hypotension prediction systems.

}


\maketitle

\section{Background}\label{sec:intro}
Hypotension in intensive care unit (ICU) patients is commonly defined as blood pressure below an intervention threshold, such as mean arterial pressure (MAP) below 65 mmHg or systolic blood pressure (SBP)  below 90 mmHg, where risk of organ damage increases \citep{Schenk2021}. Management is complex, as causes include vasodilation from sedatives, bacterial infections, or hypovolemia from fluid loss or bleeding \citep{DeBacker2022, vanderVen2022}. Timely fluid resuscitation often restores arterial pressure and organ perfusion \citep{procter2025intravenous}, but escalation to catecholamines e.g vasopressors such as norepinephrine and vasopressin or inotropic drugs like epinephrine and dobutamine is required whenever fluid resuscitation alone is not sufficient to prevent organ damage \citep{chen2025hypotensiononline}. 

There is sustained interest in using machine learning (ML) to anticipate hypotensive events in ICU patients \citep{Michard2025} Existing methods fall into two main categories: (i) predicting blood pressure decline, either by threshold breaches (e.g., MAP < 65 mmHg for 1 min) \citep{hatib2018machine, Jeong2024} or by forecasting future trajectories \citep{He2025, Kapral2024}, which support early detection but not context-specific treatment decisions; and (ii) directly predicting treatment needs, such as vasopressor use, providing a more actionable alternative however currently limited by selective populations or artificial data balancing \citep{Kwak2020, Chen2025}.

In this study, we adopt a treatment-focused prediction approach in a broad ICU cohort, explicitly accounting for the naturally imbalanced ICU monitoring data. Catecholamine initiation refers to the start of vasoactive agents used for hemodynamic support, including vasopressors (epinephrine, norepinephrine, phenylephrine, vasopressin) and inotropes (dopamine, dobutamine, milrinone). Our main contributions are: (1) defining a treatment-aligned prediction target for a heterogeneous ICU population; (2) developing and evaluating a ML model using discrete MAP measurements; (3) benchmarking against a baseline model using the last MAP value, reflecting current clinical practice; (4) performing subgroup analyses across demographic and clinical strata; and (5) integrating an explanation framework to identify influential features.

\subsection{Related works}\label{subsec:rel_work}

\subsubsection{Prediction of blood pressure decline}

Most models predict hypotension using fixed thresholds (e.g., MAP $<65$ for $\geq$1 min) \citep{Mukkamala2024}. A widely adopted example is the Hypotension Prediction Index (HPI), trained on intraoperative arterial waveform features to predict hypotension within 5-15 minutes \citep{hatib2018machine}. Its predictive performance has been validated in several studies \citep{Davies2024, Schuurmans2024}, and its utility has been investigated for specific patient subgroups \citep{PilakoutaDepaskouale2025}, though concerns remain that good performance metrics are largely driven by current MAP \citep{Tschoellitsch2025}. Beyond the HPI, many other models exist; for instance, \citet{Jeong2024} predicted intraoperative hypotension (SBP $<$90 mmHg) five minutes in advance using five non-invasive signals. Yet, fixed-threshold definitions of hypotension ignore patient-specific variation in blood pressure tolerance and so lack the necessary clinical context to differentiate how critical a given MAP value is. Another angle on the problem of predicting hypotensive events is through direct forecasting of blood pressure trajectories \citep{Pal2024}. \citet{He2025} applied a Temporal Fusion Transformer (TFT) to predict five vital signs, including MAP, using demographics, comorbidities, labs, and vasopressor use. Similarly, \citet{Kapral2024} used TFT to forecast intraoperative trajectories 7 minutes ahead, along with binary hypotension prediction, using lower-resolution physiological data. While trajectory forecasting offers insight into the expected stability of the patient, it does not guide treatment escalation. Thus, a central contribution of this work is to move beyond these paradigms towards an actionable definition.

\subsubsection{Predicting Actionable Treatment Events}

This paradigm reframes hypotension prediction as forecasting escalative therapeutic decisions, using treatment onset as a clinically meaningful marker. \citet{Kwak2020} trained bidirectional long short-term memory (bi-LSTM) models to predict vasopressor need within the first ICU day from physiological parameters (e.g., heart rate, blood pressure) in a 2-hour window. To address class imbalance, they used imputation, time-matching, and oversampling, creating a balanced dataset but deviating from the original distribution. Other studies targeted narrower cohorts: \citet{Chen2025} forecast vasopressor use, mechanical ventilation, and renal replacement therapy in patients with community-acquired pneumonia (including COVID-19), while \citet{Choi2024} predicted short-term vasopressor requirements in already hypotensive emergency department patients using deep learning on continuous arterial pressure data. Collectively, these studies highlight the promise of treatment-focused prediction but remain limited by selective populations or artificial data balancing that reduce applicability in real-world ICU settings.

\section{Methods}\label{sec:methods}

\begin{figure*}[ht!]
    \centering
    \includegraphics[width=\textwidth]{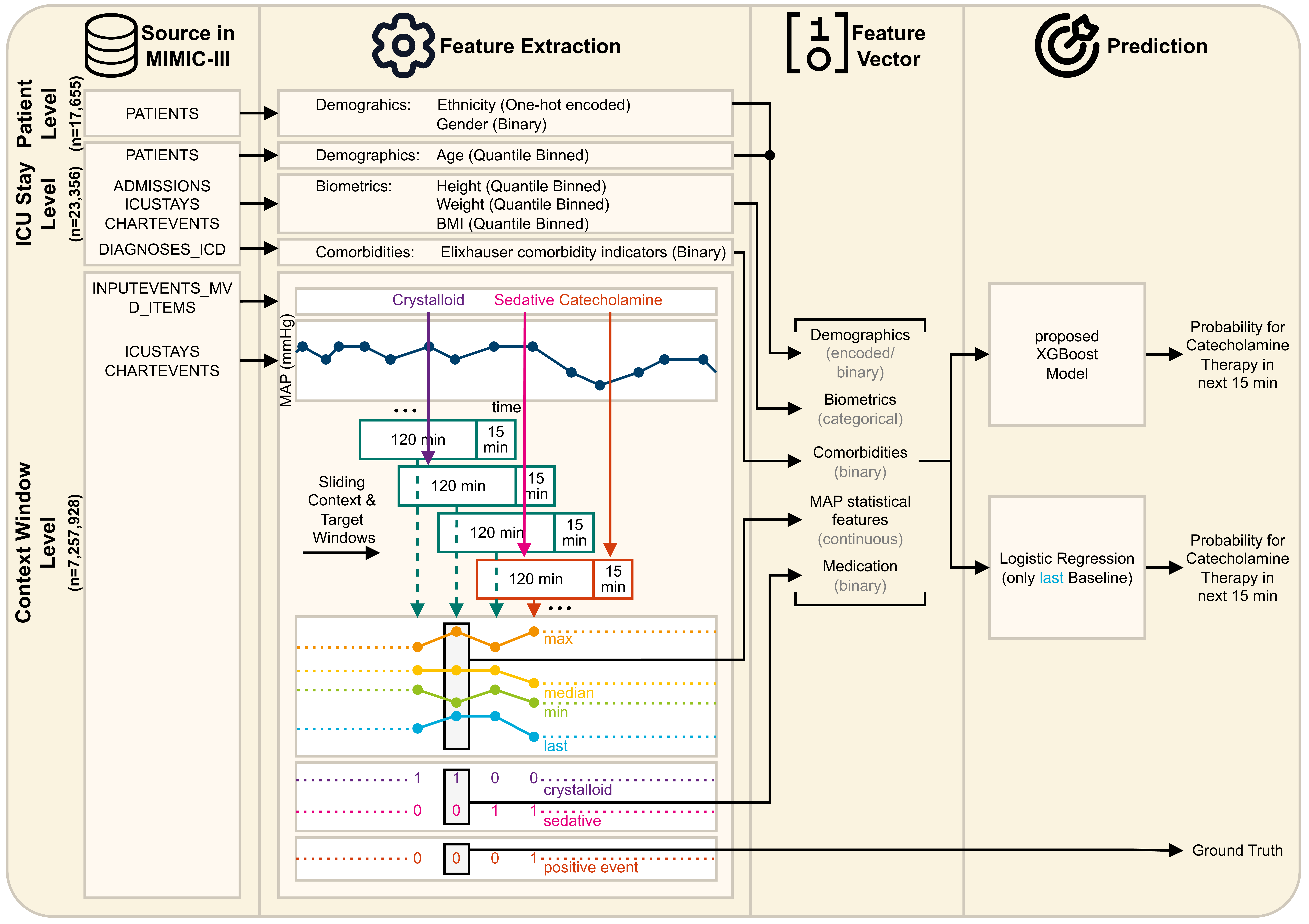}
    \caption{Overview of the MAP feature extraction and prediction workflow. Patient metadata and MAP measurements were extracted from the MIMIC-III database version 1.4. Statistical features were computed over 120-min sliding context windows with a 15-min window hop size. Binary indicators of ongoing treatments were added, and the target was defined as the occurrence of catecholamine therapy initiation within a 15-min window. An XGBoost model was compared to a baseline logistic regression using only the last MAP value in each context window which mimic current clinical practice.}
    \label{fig:data_flow}
\end{figure*}

Our methodological approach, summarized in \autoref{fig:data_flow}, covers the full workflow from dataset construction, including selection of the database source and relevant tables, through feature engineering, which incorporates data sampling at multiple time intervals, to model development and evaluation, encompassing both our proposed machine learning model and a baseline comparator.

\subsection{Dataset} \label{subsec:dataset}

We used data from the Medical Information Mart for Intensive Care database (MIMIC-III) version 1.4 \citep{MIMICiii} restricting to the Metavision system to ensure reliable start and stop times of the administered treatments. Only ICU-stays with documented MAP values in the chartevents table were included. MAP values originate from two invasive sources (item IDs 225312 and 220052) and one non-invasive source (item ID 220181), typically recorded several times per hour with variable frequency. We created a dataset by combining all available MAP values ranging from 30 to 200 mmHg with preference for invasive readings (averaging both if available). In the supplementary material, we also present results based on purely invasive and purely non-invasive measurements, c.f. Figure~\ref{app:performance_graphs} and Table~\ref{app:performance_table}.

\subsection{Prediction Task}

The goal of this study is to predict initiation of catecholamine therapy in ICU patients. Seven vasoactive agents were considered: phenylephrine, norepinephrine, epinephrine, vasopressin dopamine, dobutamine and milrinone. For each patient and visit, we defined a two-hour MAP context window to extract features, followed by a 15-minute target window with a binary label indicating a possible treatment initiation within it. To account for titration, consecutive administrations of the same drug were grouped into a single treatment episode, following the official MIMIC-III implementation \citep{mimiciii_git}. Dose adjustments were not considered new onsets, though overlapping episodes of different drugs were allowed. Therapy duration was not included. Positive labels thus correspond to the initiation of any grouped episode within the target window, while all others were negative. All context windows with at least two MAP measurements were included, preserving the natural positive-to-negative ratio. This labeling scheme defined the ground truth for model training and evaluation (Figure~\ref{fig:data_flow}).

\subsection{Features}
Our dataset comprises five feature categories. Each feature is encoded as quartile-based (Q), binary (B), categorical (C), or continuous (Cont). The first is demographics consisting of gender (B), age (Q), and ethnicity (C). The second is biometrics, containing height, weight, and body mass index (BMI) where each was encoded as quartiles (Q). The third are the comorbidities obesity, hypertension, diabetes, kidney disease, lung disease, heart disease, drug abuse, and depression, all as (B). The fourth consist of statistical features from all MAP values within the collection window, namely mean, standard deviation, median, interquartile range, minimum, maximum, first, and last recorded value, rate of change, slope, and time-weighted mean (values closer to the target window contribute more) where all of these are continuous values. The final group consists of administered treatments during the context window relevant to patient stabilization, where a list of medications was grouped into the 11 broad categories sedatives, blood products, antibiotics, anticoagulation/antiplatelets, neuromuscular blockers, analgesics, crystalloids, electrolytes, GI protection, parenteral nutrition, and antiarrhythmics, all as (B). The total set of features contains 37 features. The list of medications that are grouped into the broad 11 categories are available in the Supplementary material in Table \ref{app:medications_categories}.

\subsection{Model}\label{subsec:models}
We used Extreme Gradient Boosting (XGBoost) as the main model, chosen for its robustness to feature scaling and outliers, native handling of missing data (avoiding imputation biases), strong performance on tabular data \citep{Tabarena2025}, and compatibility with explainability frameworks. Hyperparameters were tuned with Optuna \citep{akiba2019optuna} using the Tree-structured Parzen Estimator (TPE) sampler \citep{bergstra2011algorithms}. Data were split 70:15:15 into train, validation, and test sets at patient level to prevent overlap. Training ran for up to 1000 boosting rounds with early stopping after 50 rounds without validation improvement. Predicted probabilities were calibrated using isotonic regression on the validation set to better align predicted with true outcome likelihoods. As a baseline, we implemented logistic regression using only the last MAP value in the context window (feature ''last''), mimicking current clinical practice. Full hyperparameters and search spaces are reported in the Supplementary material in Table \ref{app:hyperparameters}.

\subsection{Experimental Setup}\label{subsec:experiment_design}

All evaluations were performed on a held-out test set of patients and ICU stays excluded from training and validation. Performance metrics were grouped as follows: Discrimination was assessed using AUROC (area under the receiver operating characteristic curve), which measures overall separability. Threshold-dependent classification includes sensitivity, specificity, positive predictive value (PPV), and negative predictive value (NPV). Calibration includes calibration slope and intercept, Brier score, and expected calibration error (ECE). Ninety-five percent confidence intervals were estimated via 1,000 empirical bootstrap iterations. Thresholded metrics were evaluated at a single operating point chosen to achieve test sensitivity of at least 0.80, controlling for sensitivity while highlighting trade-offs in specificity, PPV, and calibration, for additional decision thresholds see the Supplementary Material in Table \ref{app:decision_thresholds}. To analyze model performance at a clinically relevant sensitivity range [0.75-0.85], Deep ROC analysis was performed, which reports AUROC values restricted to this range rather than across the full curve \citep{Carrington2023}. Subgroup analyses were conducted to explore potential performance differences across patient demographics, biometrics, comorbidities, medication exposures, and MAP-related strata. Model interpretability was addressed using the SHAP (Shapley Additive Explanations) treeexplainer, which provides exact Shapley values for trees \citep{Lundberg2020}.

\section{Results}\label{sec:results}

\subsection{Descriptive statistics} \label{subsec:descr_stats}

\begin{table}[ht!]
\centering
\tiny
\setlength{\tabcolsep}{1pt}
\caption{Descriptive statistics of the dataset at three levels: (1) Patient-level characteristics including gender and ethnicity distributions; (2) ICU-stay-level characteristics, showing frequency of catecholamine therapy onsets by drug and the fraction of ICU stays with prescriptions for the respective drug on the ICU admission day (proxy for pre-ICU exposure). Further differences in demographics, comorbidities, and concurrent treatments between positive (vasoactive therapy onset in target window) and negative stays;  (3) Context-window-level characteristics, reporting prevalence and distribution of MAP measurements for positive and negative stays.}

\label{tab:descriptive_stats}
\begin{tabular}{@{}lllrr@{}}
\toprule
\multicolumn{3}{l}{\textbf{Patient-Level Characteristics} (n = 17,655 patients)} & \multicolumn{2}{c}{\textbf{Overall  (\%)}} \\
\midrule
Demographics & Gender & Male / Female & \multicolumn{2}{c}{44 / 56} \\
            & Ethnicity & White / Black / Hispanic & \multicolumn{2}{c}{72.8 / 10.6 / 4.0} \\
             &           & Asian / Other            & \multicolumn{2}{c}{2.6 / 9.9} \\
\midrule

\multicolumn{3}{l}{\textbf{ICU-Level Characteristics} (n = 23,356 stays)} & \textbf{ICU (\%)} & \textbf{Pre-ICU (\%)} \\
\midrule
\makecell[l]{Catecholamine Onset} & Any & \% of all stays & 30.4 & 15.5 \\
& Phenylephrine   & \% of positive stays & 68.3 & 38.9 \\
&Norepinephrine  & \% of positive stays & 45.1 & 13.7 \\
&Epinephrine     & \% of positive stays & 10.4 & 5.7 \\
&Dopamine        & \% of positive stays & 12.8 & 0 \\
&Dobutamine      & \% of positive stays & 2.9 & 0 \\
&Milrinone       & \% of positive stays & 6.3 & 3.7 \\
&Vasopressin     & \% of positive stays & 14.1 & 5.4 \\
\midrule
\multicolumn{3}{l}{\textit{\textbf{By Outcome}}} & \textbf{Positive} & \textbf{Negative} \\
ICU-Stay         & Length of stay & days, Median [IQR] & 3.3 [1.8-7.2] & 1.8 [1.1-2.9] \\
Demographics & Age & years, Median [IQR] & 68 [58-78] & 64 [52-78] \\
Biometrics  & BMI & kg/m$^2$, Median [IQR] & 27.6 [24.1-31.8] & 27.0 [23.4-31.6] \\
Comorbidities & Hypertension & \% & 64.6 & 57.9 \\
             & Heart disease & \% & 40.9 & 28.0 \\
             & Diabetes & \% & 31.7 & 27.0 \\
Treatment & Crystalloids & \% & 100.0\% & 95.1\% \\
            & Antibiotics & \% & 87.5\% & 53.5\% \\
            & Electrolytes & \% & 88.9\% & 57.5\% \\
            & Analgesics & \% & 82.0\% & 46.2\% \\
            & Sedatives & \% & 80.0\% & 36.2\% \\
            & Blood products & \% & 58.9\% & 24.6\% \\
            & Neuromusc. block. & \% & 8.8\% & 0.9\% \\
\midrule
\multicolumn{3}{l}{\textbf{Context-Window Characteristics} (n = 7,257,928 windows)} & \textbf{Positive} & \textbf{Negative} \\
\midrule
Counts      & Number of windows & n (\%) & 13,364 (0.1841) & 7,244,564 (99.8159) \\
            & MAP Count / Window & count, Median [IQR] & 3.0 [2.0-4.0] & 2.0 [2.0-2.0] \\
MAP Trends  & Average MAP & mmHg, Median [IQR] & 69.0 [62.3-77.4] & 76.5 [68.0-87.0] \\
            & Lowest MAP  & mmHg, Median [IQR] & 62.0 [54.0-70.0] & 72.0 [64.0-82.0] \\
            & Std MAP     & mmHg, Median [IQR] & 6.4 [3.2-11.2] & 4.2 [2.1-8.5] \\
            & Last MAP    & mmHg, Median [IQR] & 67.0 [59.0-77.0] & 76.0 [67.0-87.0] \\
            & MAP Slope   & mmHg/s, Mean $\pm$ SD & -0.0009 $\pm$ 0.0159 & -0.0001 $\pm$ 0.0067 \\
\bottomrule
\end{tabular}
\end{table}

At the patient level ($n=17{,}655$), the cohort was demographically imbalanced, with 72.8\% white patients and underrepresentation of black (10.6\%), hispanic (4.0\%), and asian (2.6\%) patients; the gender distribution was moderately balanced (44\% male vs.\ 56\% female). At the ICU-stay level ($n=23{,}356$), catecholamine therapy was initiated in 30\% of stays, predominantly phenylephrine and norepinephrine, while dopamine, dobutamine, and milrinone were rare (<7\%). Therapy had already been prescribed on the ICU admission day in 15.5\% of positive stays, mainly for phenylephrine. Positive ICU stays were older, longer, had higher prevalence of cardiovascular comorbidities, and received more adjunctive treatments (e.g., antibiotics, sedatives, blood products, neuromuscular blockade), reflecting higher acuity. At the context-window level ($n=7.26$M), only 0.18\% of all windows were positive, highlighting the extreme class imbalance. Importantly, no measures were taken to balance this imbalance, e.g., through over- or undersampling as it represents the true underlying event distribution in the dataset. Positive windows were preceded by lower and more variable MAP values, slightly more frequent measurements, and a downward trend, consistent with hemodynamic instability. Full statistics, including MAP distributions and drug-specific details, are provided in Table~\ref{tab:descriptive_stats}.

\subsection{Label Alignment} \label{subsec:lab_align} 

\begin{figure*}[ht!]
    \centering
    \includegraphics[width=\textwidth]{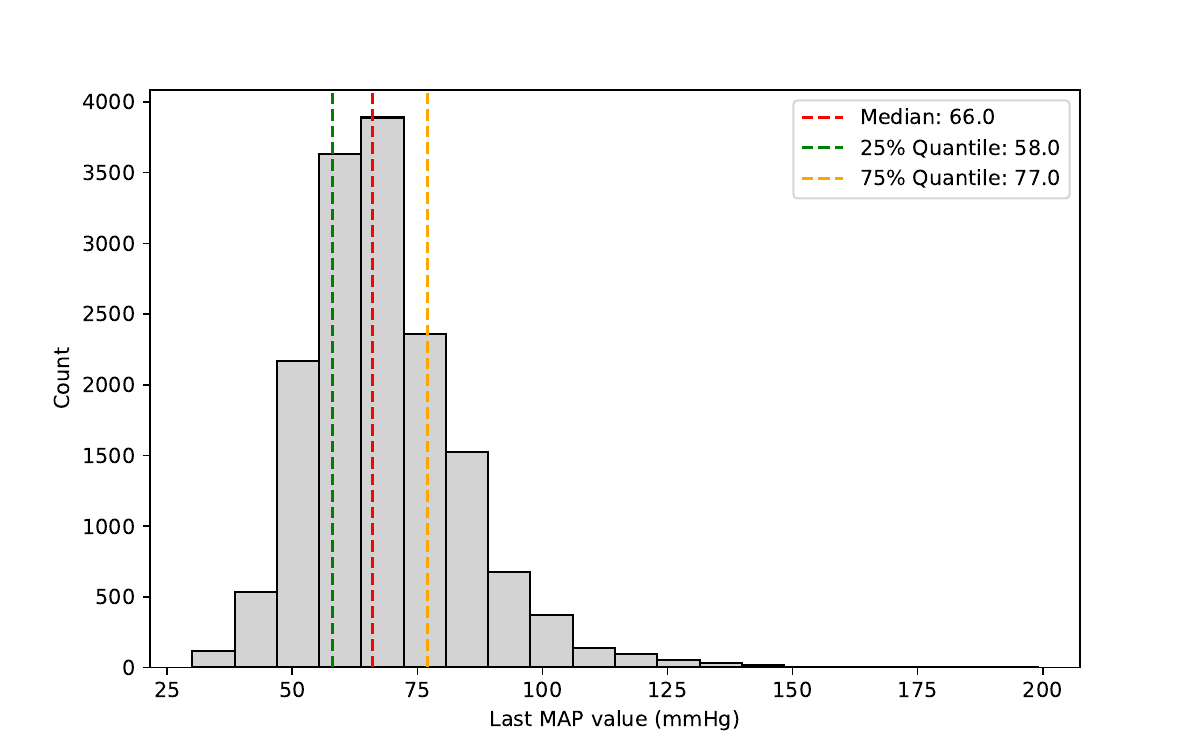}
    \caption{Distribution of the last recorded MAP value preceding catecholamine treatment onset. Last measurements were taken from the target window but preceding treatment start if available, otherwise from the context window. Vertical lines indicate the median (red dashed), 25th percentile (green dashed), and 75th percentile (orange dashed).}

    \label{fig:label_alignment}
\end{figure*}

We analyzed the last MAP measurements preceding escalation of catecholamine therapy. Figure~\ref{fig:label_alignment} shows that escalation of hypotension management often occurs at MAP values above the conventional hypotension threshold of 65 mmHg, suggesting that factors beyond absolute MAP - such as trends, patient history, or clinician judgment - influence treatment decisions at bedside.

\FloatBarrier

\subsection{Predictive performance}
\FloatBarrier
\begin{figure*}[ht]
    \centering
    \includegraphics[width=\textwidth]{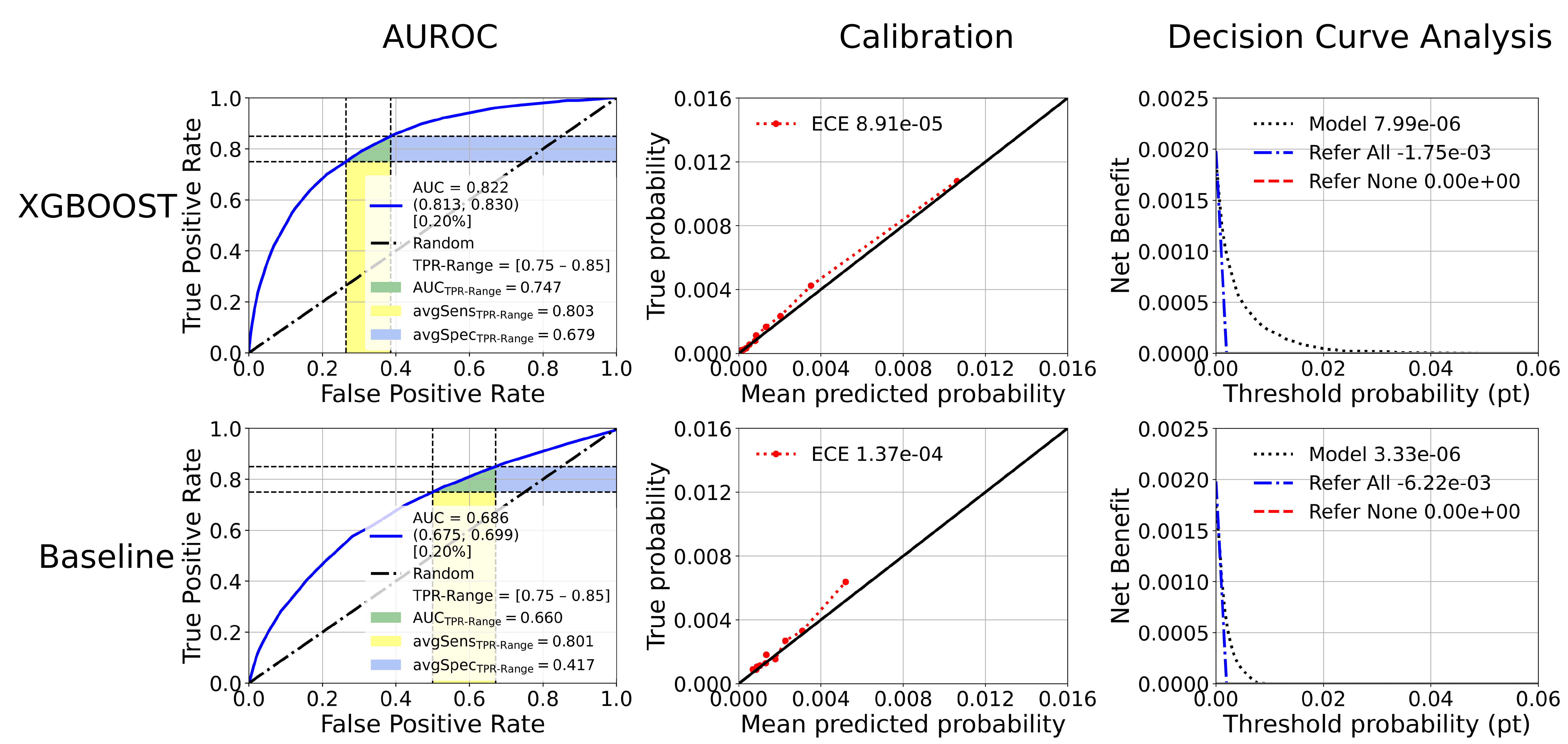}
    \caption{AUROC (Confidence-Interval 95 \%), calibration, and net benefit curves for the XGBoost and baseline models}
    \label{fig:performance}
\end{figure*}

Overall performance was evaluated using AUROC, calibration, and decision curve analysis (Figure~\ref{fig:performance}). Compared to the baseline, XGBoost achieved higher AUROC (0.822, 95\% CI: 0.813-0.830 vs. 0.686, 95\% CI: 0.675-0.699), higher partial AUROC in the clinically relevant sensitivity range (0.747 vs. 0.660), lower calibration error (ECE 8.91e-05 vs. 1.31e-04), and the greatest net benefit across thresholds.

\FloatBarrier
\begin{table}[ht!]
\centering
\caption{Additional performance metrics for XGBoost and Baseline models, using thresholds selected to target a test sensitivity of at least 80\%.}
\label{tab:performance_metrics_transposed}
\small
\begin{tabular}{lcc}
\toprule
\textbf{Metric} & \textbf{XGBoost} & \textbf{Baseline} \\
\midrule
     Specificity (CI) $\uparrow$ & \textbf{0.6118 (0.6109, 0.6127)} & 0.3662 (0.3654, 0.3669) \\
             Positive Predictive Value (CI) $\uparrow$ & \textbf{0.0043 (0.0041, 0.0045)} & 0.0026 (0.0025, 0.0027) \\
             Negative Predictive Value (CI) $\uparrow$ & \textbf{0.9995 (0.9995, 0.9996)} & 0.9991 (0.9990, 0.9992) \\
                           Prevalence & 0.20\% & 0.20\% \\
\bottomrule
\end{tabular}%
\end{table}

To complement the curve-based evaluation, Table~\ref{tab:performance_metrics_transposed} reports additional performance metrics at the test-set threshold selected to achieve 0.80 sensitivity. At this operating point, and given the low event prevalence of 0.2\%, the XGBoost model outperformed the baseline across all measures , with higher specificity (0.612 vs. 0.366), higher positive predictive Value (0.0043 vs. 0.0026) and higher negative predictive value (0.9995 vs. 0.9991).

\subsection{Performance across patient subgroups}\label{subsec:subgroup_performance}

\begin{table}[ht!]
\setlength{\tabcolsep}{.8pt}
\caption{Subgroup performance across various input features. Reported are AUROC, sensitivity, specificity, and event prevalence [pos/total], with thresholds targeting $\geq 80\%$ sensitivity on the test set. Only subgroups with AUROC above the overall model performance of 0.822 are included.}
\begin{tabular}{p{0.3\textwidth} c c c c}
\toprule
Feature & AUROC $\uparrow$ & Sens $\uparrow$ & Spec $\uparrow$ & [pos/total]  \\
\midrule
gender: Male & 0.829 & 0.858 & 0.624 & [1349/660K] \\
ethnicity: White & 0.825 & 0.861 & 0.611 & [1579/831K] \\
ethnicity: Black & 0.830 & 0.880 & 0.611 & [217/94.7K]  \\
age $<$ 53.0 & 0.830 & 0.796 & 0.714 & [304/250K] \\
height [170.5 - 178.0) & 0.826 & 0.876 & 0.556 & [307/127K] \\
height $>$= 178.0 & 0.833 & 0.837 & 0.675 & [1007/581K] \\
weight [65.6 - 79.6) & 0.828 & 0.869 & 0.602 & [542/285K] \\
weight $>$= 93.6 & 0.826 & 0.864 & 0.602 & [617/316K] \\
bmi $>$= 32.1 & 0.834 & 0.838 & 0.664 & [904/566K] \\
obesity = 0 & 0.824 & 0.854 & 0.616 & [2024/1011K] \\
hypertension = 0 & 0.823 & 0.847 & 0.621 & [822/464K]  \\
diabetes = 0 & 0.827 & 0.851 & 0.628 & [1459/776K] \\
kidney disease = 0 & 0.826 & 0.851 & 0.629 & [1648/871K] \\
lung disease = 1 & 0.822 & 0.865 & 0.579 & [770/344K] \\
heart disease = 0 & 0.822 & 0.823 & 0.662 & [1295/727K] \\
drug abuse = 0 & 0.822 & 0.857 & 0.607 & [2133/1060K] \\
depression = 1 & 0.832 & 0.841 & 0.659 & [308/174K] \\
blood products  = 0 & 0.822 & 0.839 & 0.630 & [1888/1048K] \\
antibiotics  = 0 & 0.831 & 0.851 & 0.634 & [1687/910K]  \\
antibiotics  = 1 & 0.779 & 0.857 & 0.516 & [525/206K]  \\
anticoag/antiPLT = 0 & 0.825 & 0.854 & 0.619 & [1784/905K] \\
analgesics = 0 & 0.825 & 0.783 & 0.704 & [1203/822K]\\
crystalloids = 0 & 0.903 & 0.733 & 0.916 & [101/137K]  \\
electrolytes = 0 & 0.830 & 0.834 & 0.655 & [1466/866K]\\
GI protection  = 0 & 0.825 & 0.855 & 0.615 & [2045/1037K] \\
parenteral nutrition = 0 & 0.823 & 0.853 & 0.614 & [2054/1039K] \\
antiarrhythmics = 0 & 0.823 & 0.842 & 0.626 & [2012/1069K] \\
last $\leq 65$ & 0.780 & 0.967 & 0.207 & [1046/229K]\\
65 < last $\leq 70 $& 0.759 & 0.874 & 0.463 & [333/147K]\\
70 < last $\leq 100 $ & 0.803 & 0.707 & 0.758 & [737/650K]\\
\bottomrule
\end{tabular}
\label{tab:subgroup_performance}
\end{table}

The subgroup analysis (Table~\ref{tab:subgroup_performance}) shows performance for subgroups with AUROC above the overall test-set performance of 0.822. Across demographics, biometrics, and comorbidities, AUROC was generally consistent (0.822-0.834), with subgroups with or without conditions such as obesity or hypertension showing similar performance. For most additional treatments, AUROC was lower in windows where the medication had already been administered, which also had higher event prevalence. The largest decreases were observed for blood products, sedatives, and analgesics. Stratification by last MAP value reflected the distributions in Table~\ref{tab:descriptive_stats}: windows with MAP $\leq 65$ mmHg had near-perfect sensitivity but very low specificity, 70-100 mmHg showed the most balanced trade-off. Full subgroup results are provided in Supplementary Table ~\ref{app:subgroup_performance}.

\subsection{Interpretability}\label{subsec:interpretability}

\begin{figure*}[ht!]
    \centering
    \includegraphics[width=0.8\textwidth]{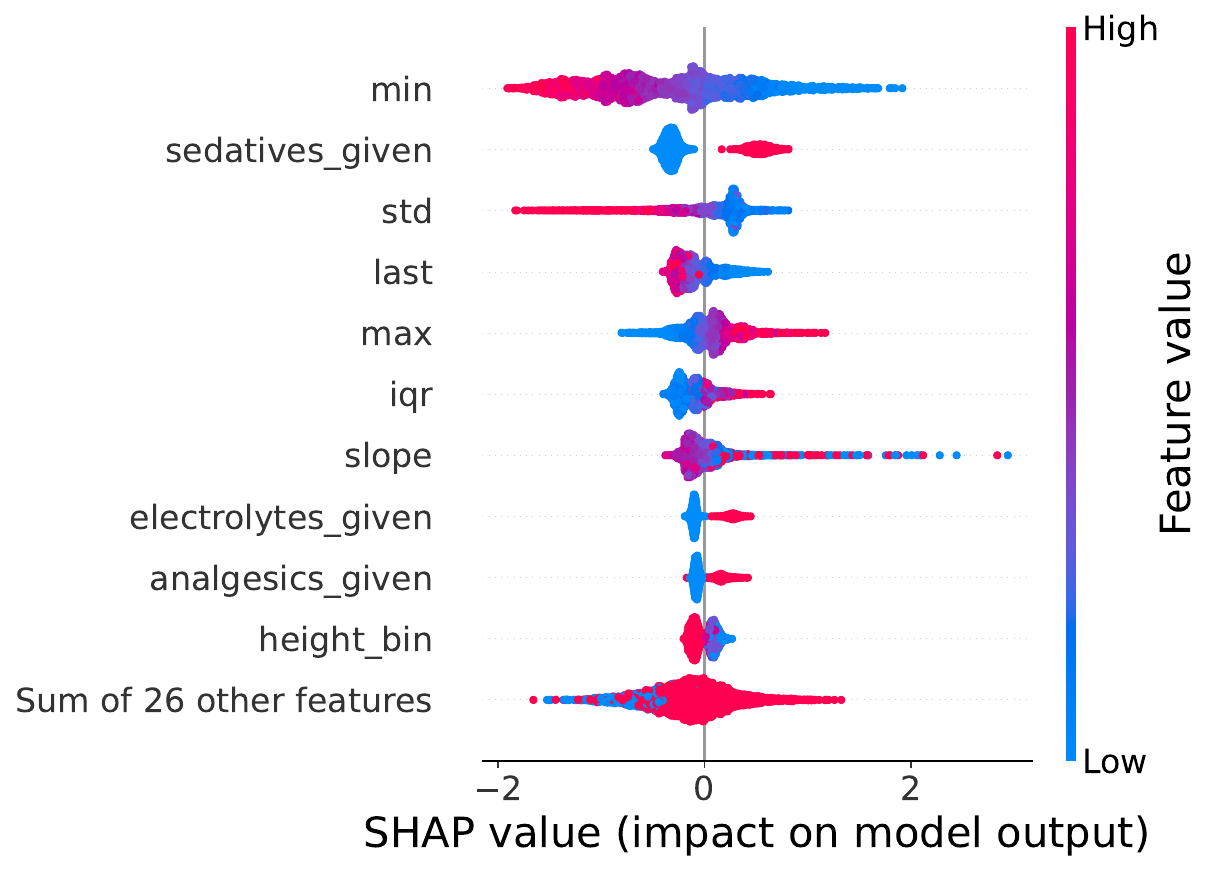}
    \caption{Ten most influential features according to global SHAP values ranked by mean absolute SHAP value}
    \label{fig:shap_importance}
\end{figure*}

Figure~\ref{fig:shap_importance} shows the ten most influential features based on mean absolute SHAP values, with color encoding feature value (red = high, blue = low, purple = intermediate). \textbf{Blue (lower feature values):} 
Lower minimum and last MAPs in the context window were associated with higher predicted risk. Absence of treatments such as sedatives, electrolytes, and analgesics also reduced predicted risk. Low standard deviation contributed positively, while low maximum values contributed negatively.

\textbf{Red (higher feature values):} Higher minimum MAPs and higher MAP standard deviation generally reduced risk, whereas higher MAP IQR, maximum values, and treatment administration increased predicted risk.

\textbf{Purple (intermediate values):} Intermediate feature values were mostly neutral. Near-zero MAP slopes indicated stable MAP and lower risk, while extreme slopes increased predicted risk. \newline Demographics, biometrics, and comorbidities-such as age, height, BMI, and obesity-ranked lower and contributed weakly. Full feature importance across all features is shown in Supplementary Figures ~\ref{app:shap_mix_full} to ~\ref{app:shap_noninv_full}.

\section{Discussion}\label{sec:discussion}

\subsection{Clinical utility}

Our results highlight the potential of predicting catecholamine therapy initiation as a clinically actionable event signaling the necessity to escalate hypotension management.
This framing enables predictive systems that integrate physiological trends with contextual patient information, moving beyond MAP-based alarms toward trustworthy, context-aware decision support. Such models could serve as adjunct decision-support tools in the ICU, alerting clinicians to patients at elevated short-term need of catecholamine therapy thus allowing earlier preparation for escalation of care. 

\subsection{Subgroup performance and fairness considerations} 

Subgroup analyses indicated broadly stable discrimination across demographic, biometric, and clinical categories, suggesting that no subgroup is systematically disadvantaged. Lower discrimination was observed in subgroups already receiving certain medications, such as sedatives, blood products, neuromuscular blockers, and analgesics. This likely reflects higher illness severity and complex physiological patterns that mask hypotensive trends. Windows with low MAP values (e.g., last MAP $\leq 65\,\text{mmHg}$) drove high sensitivity but also elevated false positive rates, indicating that low MAP strongly signals imminent hypotension but increases false alarms under intensive monitoring.

\subsection{Feature importance} 

Our SHAP analysis underscores that recent MAP dynamics and concurrent treatment indicators are the main drivers of predicted catecholamine initiation, whereas static patient attributes and comorbidities contribute minimally. This aligns with subgroup analyses showing consistent discrimination across demographics and comorbidity profiles, suggesting the model primarily leverages acute physiological signals rather than static risk factors.

The minimum MAP feature was the single most influential predictor, affecting both positive predictions (when low) and negative predictions (when high). Beyond low MAP values, the model is most sensitive to features capturing MAP dynamics, including slopes, variability, and interquartile range, as well as concurrent treatments. This highlights that predictions are driven not solely by extreme hypotension, but by the pattern and evolution of MAP over time, consistent with prior findings on hypotension prediction \citep{Mulder2024,Tschoellitsch2025}.

Treatment-related features-such as administered sedatives or electrolytes also ranked highly, reflecting their prevalence in positive cases and serving as indicators of active interventions and patient deterioration, reinforcing that the model integrates both physiological signals and the clinical context to make actionable predictions.

\subsection{Limitations}

The primary limitation of our current model is the relatively low specificity across different patient groups which stems in part from the highly imbalanced dataset. Even high-performing models generate many false positives in this setting, limiting positive predictive value \citep{vickers2006decision,tenny2025prevalence}. This in turn limits clinical applicability by contribute to alarm fatigue in practice. A promising direction to deal with this is to consider the uncertainty of the prediction for example through conformal prediction \citep{Shashikumar2021}. A second limitation is limited temporal awareness: Each prediction window is treated in isolation, without mechanisms to reset risk after interventions or distinguish new deterioration from residual signals. This is further compounded by using discrete, intermittently charted MAP measurements rather than continuous waveforms, which may miss rapid physiological changes. This limitation might be partially mitigated by more complex models that aggregate information over longer temporal contexts, for example via a hidden state \citep{Ohno2021}. Finally, the available context variables are relatively narrow. Additional attributes, such as ICU stay duration, ICU type, or broader clinical context, could help the model interpret physiological changes and treatment patterns more effectively.

\subsection{Future Work}
Future work could make use of continuous ABP waveforms instead of discrete MAP readings, incorporate additional vital signs (e.g., heart rate, ECG, body temperature), and model dosage trajectories of concurrent medications (e.g., sedatives, blood products, analgesics). In addition, future research should aim to benchmark this treatment-focused prediction approach against existing hypotension prediction systems and blood pressure forecasting models. Such comparisons, ideally conducted on shared datasets or within standardized evaluation benchmarks, would help contextualize the performance of our model, despite differences in target definitions, and clarify its relative strengths and limitations.

\section{Conclusions}\label{sec:conclusions}

This study demonstrates that predicting catecholamine therapy initiation in a general ICU population is feasible and provides a clinically actionable complement to fixed-threshold definitions of hypotension. By evaluating all available context windows, our model captures the natural prevalence and distribution of hypotension management escalations, reflecting realistic clinical practice. Fixed-threshold MAP ranges alone miss a substantial proportion of positive events, highlighting the value of treatment-aligned prediction. The XGBoost model consistently outperformed a logistic regression baseline using only the last recorded MAP, providing interpretable predictions across demographic and clinical subgroups. Performance was robust in clinically relevant sensitivity ranges, though low overall event prevalence remains a challenge for rare-event prediction. All code used for data processing, model training, and evaluation is available to ensure reproducibility of the presented results. Future work should focus on enriching input features and benchmarking against established hypotension prediction systems.

\section*{List of abbreviations}

ICU: Intensive care unit. MAP: Mean arterial blood pressure. mmHg: Millimetres of mercury.
ML: Machine learning. HPI: Hypotension Prediction Index. TFT: Temporal fusion transformer. Auto-ML: Automatic Machine learning. TPE: Tree-structured parzen estimator. XGBoost: Extreme gradient boosting. AUROC: Area under the decision operator curve. AUC: Area under curve. PPV: Positive predictive value. NPV: Negative predictive value. AUPRC: Area under the precision recall curve. ECE: Expected calibration error. BMI: Body mass index. SHAP: Shapley additive explanations. bi-LSTM: bidirectional long short-term memory. ECG: Electrocardiogram.

\section*{Declarations}\label{sec:declaration}

\subsection{Ethics approval and consent to participate}
This study utilized the publicly available MIMIC-III database, which comprises de-identified health data from patients admitted to the Beth Israel Deaconess Medical Center. The database was approved by the Institutional Review Boards (IRBs) at both the Massachusetts Institute of Technology (MIT) and Beth Israel Deaconess Medical Center. Given that the data is fully de-identified and the study was retrospective in nature, the requirement for individual patient consent was waived by the IRBs.

\subsection{Consent for publication}
Not applicable

\subsection{Availability of data and materials}
Code for dataset preprocessing and experimental replications can be found in our dedicated repository \url{https://github.com/AI4HealthUOL/actionable-hypotension}.

\subsection{Competing interests}
Upon manuscript submission, all authors completed the author disclosure form, confirming the absence of any conflicts of interest.

\subsection{Funding}
This research received no specific grant from any funding agency in the public, commercial, or not-for-profit sectors.

\subsection{Authors' contributions}
Conceptualization: N. Saibel, R. Koebe, J.M. Lopez Alcaraz, N. Strodthoff.
Methodology: N. Saibel, R. Koebe.
Software: N. Saibel, R. Koebe.
Validation: N. Saibel, R. Koebe.
Formal Analysis: N. Saibel, R. Koebe.
Investigation: N. Saibel, R. Koebe.
Resources: J.M. Lopez Alcaraz, N. Strodthoff.
Data Curation: N. Saibel, R. Koebe.
Writing - Original Draft: N. Saibel, R. Koebe.
Writing - Review \& Editing: N. Saibel, R. Koebe, S. Schäfer, J.M. Lopez Alcaraz, N. Strodthoff.
Visualization: N. Saibel, R. Koebe.
Supervision: J.M. Lopez Alcaraz, S. Schäfer, N. Strodthoff.
Project Administration: J.M. Lopez Alcaraz, S. Schäfer, N. Strodthoff.

\subsection{Acknowledgements}
Not applicable

\bibliography{sn-bibliography} 

\newpage
\begin{appendices}

\renewcommand{\thetable}{S\arabic{table}}  
\setcounter{table}{0}  

\renewcommand{\thefigure}{S\arabic{figure}}
\setcounter{figure}{0}

\section{Medication categories}

\begin{table}[ht!] 
\centering
\caption{Individual treatments categorized into 11 broad treatment classes. The table lists the specific medication labels that were mapped to each class based on their item descriptions in the MIMIC-III Metavision inputevents.}
\label{app:medications_categories}
\begin{tabular}{p{0.25\textwidth} p{0.7\textwidth}}
\hline
\textbf{Category} & \textbf{Medications / Labels} \\
\hline
Crystalloids & NaCl 0.9\%, Dextrose 5\%, Free Water, LR, D5NS, D5 1/2NS, D5LR, D5 1/4NS, NaCl 0.45\%, Sterile Water, Dextrose 10-50\%, NaCl 3\%, NaCl 23.4\% \\
Electrolytes & Potassium Chloride, K Phos, Na Phos, Calcium Gluconate, Calcium Chloride, Magnesium Sulfate, Sodium Bicarbonate 8.4\%, Hydrochloric Acid \\
Antibiotics & Cefepime, Vancomycin, Ceftriaxone, Levofloxacin, Azithromycin, Metronidazole, Bactrim, Cefazolin, Ciprofloxacin, Meropenem, Piperacillin/Tazobactam, Tobramycin, Doxycycline, Linezolid, Daptomycin, Ampicillin/Sulbactam, Acyclovir, Clindamycin, Aztreonam, Colistin, Amikacin, Imipenem/Cilastatin, Ceftaroline, Rifampin, Erythromycin, Gentamicin, Nafcillin, Tamiflu, Penicillin G, Keflex, Quinine, Isoniazid, Ethambutol, Pyrazinamide \\
Vasopressors & Epinephrine, Norepinephrine, Vasopressin, Dobutamine, Dopamine, Phenylephrine, Isuprel, Angiotensin II \\
Inotropes & Epinephrine, Dobutamine, Dopamine, Isuprel, Milrinone \\
Antiarrhythmics & Amiodarone, Esmolol, Lidocaine, Procainamide, Verapamil, Diltiazem, Adenosine \\
Anticoagulants / Antiplatelets & Heparin, Enoxaparin, Bivalirudin, Eptifibatide, Coumadin, Argatroban, Fondaparinux, Tirofiban, Abciximab, Lepirudin, Citrate, Protamine sulfate \\
Sedatives & Propofol, Midazolam, Lorazepam, Diazepam, Dexmedetomidine, Ketamine, Pentobarbital \\
Analgesics & Fentanyl, Morphine Sulfate, Hydromorphone, Meperidine, Acetaminophen-IV, Methadone, Ketorolac, Naloxone \\
Neuromuscular Blockers & Vecuronium, Rocuronium, Cisatracurium, Neostigmine \\
GI Protection & Ranitidine, Pantoprazole, Famotidine, Lansoprazole, Omeprazole, Sucralfate, Esomeprazole \\
Blood Products / Transfusions & pRBCs, Platelets, FFP, Cryoprecipitate, Whole Blood, Albumin, IVIG, Factor VIII, Factor IX, PCC, Recombinant Factor VIIa, Fibrinogen Concentrate, Thrombin, TXA, Erythropoietin, Iron Sucrose, Iron Dextran, Iron Gluconate, Ferumoxytol \\
Parenteral Nutrition & TPN w/ or w/o Lipids, Peripheral Parenteral Nutrition, Dextrose PN, Amino Acids, Lipids 10-20\%, Lipid additives \\
\hline
\end{tabular}
\end{table}

\newpage

\section{Hyperparameter}

\begin{table}[ht!]
\centering
\normalsize
\setlength{\tabcolsep}{1pt}
\caption{Hyperparameter optimization was performed using Optuna \citep{akiba2019optuna} with the TPE (Tree-structured Parzen Estimator) sampler \citep{bergstra2011algorithms} on the mixed model dataset. The resulting parameters were then applied to both the invasive and non-invasive datasets.}
\label{app:hyperparameters}
\begin{tabular}{lcc}
\hline
\textbf{Hyperparameter} & \textbf{Final Value} & \textbf{Search Space} \\
\hline
learning rate         & 0.0355      & [0.005, 0.3] (log scale) \\
max depth             & 7           & [3, 15] \\
min child weight     & 13          & [1, 20] \\
gamma                  & 2.7709      & [0, 10] \\
subsample              & 0.7696      & [0.3, 1.0] \\
colsample bytree      & 0.8996      & [0.3, 1.0] \\
colsample bylevel     & 0.9287      & [0.3, 1.0] \\
colsample bynode      & 0.8521      & [0.3, 1.0] \\
reg alpha             & 0.0319      & [1e-8, 10.0] (log scale) \\
reg lambda            & 3.57e-7     & [1e-8, 10.0] (log scale) \\
max delta step       & 3           & [0, 10] \\
\hline
\end{tabular}
\label{app:optuna_hyperparameters}
\end{table}

\newpage

\section{Performance of All Models}\label{app:three_performances}

\begin{figure}[ht!]  
\centering
\includegraphics[width=0.8\textwidth, height=0.8\textheight, keepaspectratio]{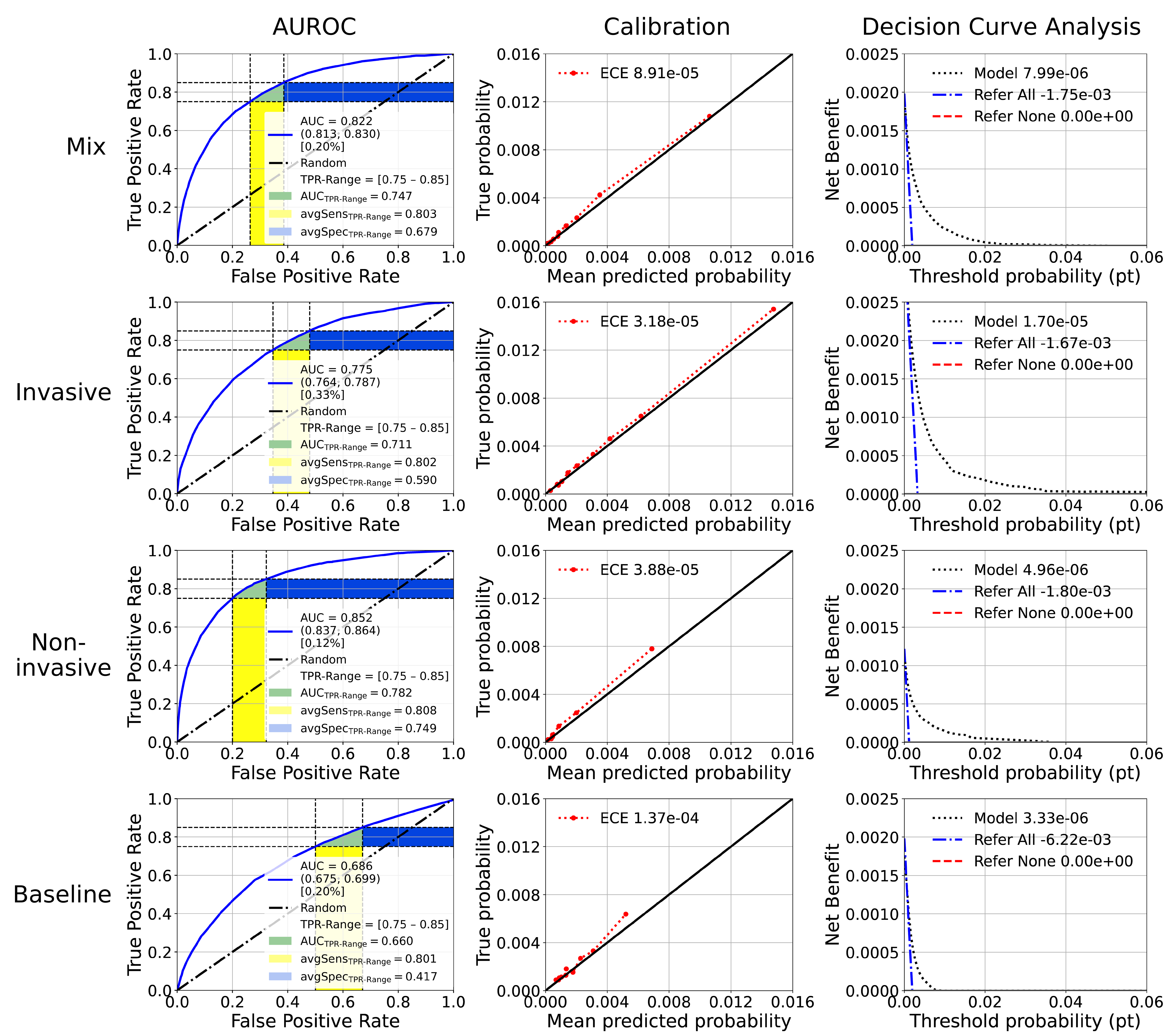}
\caption{AUROC (Confidence-Interval 95 \%), calibration, and net benefit curves for the mix, invasive, non-invasive, and baseline models.}
\label{app:performance_graphs}
\end{figure}

\begin{table}
\centering
\tiny
\caption{Performance metrics for each model, using thresholds selected to target a test sensitivity of at least 80\%.}
\label{app:performance_table}
\begin{tabular}{lcccc}
\toprule
                      \textbf{Metric} &            \textbf{Mix} &       \textbf{Invasive} &   \textbf{Non-invasive} &       \textbf{Baseline} \\
\midrule
     Specificity (95\% CI) $\uparrow$ & 0.6118 (0.6109, 0.6127) & 0.5578 (0.5563, 0.5591) & \textbf{0.7468 (0.7457, 0.7478)} & 0.3662 (0.3654, 0.3669) \\
             PPV (95\% CI) $\uparrow$ & 0.0043 (0.0041, 0.0045) & \textbf{0.0061 (0.0058, 0.0064)} & 0.0039 (0.0036, 0.0042) & 0.0026 (0.0025, 0.0027) \\
             NPV (95\% CI) $\uparrow$ & 0.9995 (0.9995, 0.9996) & 0.9990 (0.9988, 0.9991) & \textbf{0.9997 (0.9996, 0.9997)} & 0.9991 (0.9990, 0.9992) \\
             Brier Score $\downarrow$ &                  0.0020 &                  0.0033 &                  \textbf{0.0012} &                  0.0020 \\
                           Prevalence &                  0.20\% &                  \textbf{0.33}\% &                  0.12\% &                  0.20\% \\
\bottomrule

\end{tabular}
\end{table}
\FloatBarrier

\section{Decision Thresholds}

\begin{table}[ht!]
\centering
\normalsize
\setlength{\tabcolsep}{3pt}
\caption{Recommended decision thresholds from the test set for predefined target sensitivities. 
Identical thresholds may occur due to the discrete nature of model scores.}
\begin{tabular}{lcccccc}
\toprule
\multirow{2}{*}{\textbf{Metric / Result}} 
& \multicolumn{6}{c}{\textbf{Threshold}} \\
\cmidrule(lr){2-7}
& 0.70 & 0.75 & 0.80 & 0.85 & 0.90 & Youden \\
\midrule
\multicolumn{7}{l}{\textbf{Mix}} \\
Sens $\uparrow$ & 0.703 & 0.791 & 0.852 & 0.852 & 0.917 & 0.703 \\
Spec $\uparrow$ & 0.787 & 0.694 & 0.612 & 0.612 & 0.482 & 0.787 \\
PPV $\uparrow$  & 0.0065 & 0.0051 & 0.0043 & 0.0043 & 0.0035 & 0.0065 \\
Thr             & 0.00223 & 0.00161 & 0.00115 & 0.00115 & 0.00083 & 0.00223 \\
\midrule
\multicolumn{7}{l}{\textbf{Invasive}} \\
Sens $\uparrow$ & 0.731 & 0.752 & 0.825 & 0.856 & 0.906 & 0.731 \\
Spec $\uparrow$ & 0.678 & 0.652 & 0.558 & 0.512 & 0.420 & 0.678 \\
PPV $\uparrow$  & 0.0074 & 0.0071 & 0.0061 & 0.0058 & 0.0051 & 0.0074 \\
Thr             & 0.00325 & 0.00293 & 0.00210 & 0.00195 & 0.00138 & 0.00325 \\
\midrule
\multicolumn{7}{l}{\textbf{Non-invasive}} \\
Sens $\uparrow$ & 0.769 & 0.769 & 0.807 & 0.887 & 0.923 & 0.769 \\
Spec $\uparrow$ & 0.786 & 0.786 & 0.747 & 0.607 & 0.513 & 0.786 \\
PPV $\uparrow$  & 0.0044 & 0.0044 & 0.0039 & 0.0027 & 0.0023 & 0.0044 \\
Thr             & 0.00098 & 0.00098 & 0.00081 & 0.00046 & 0.00046 & 0.00098 \\
\midrule
\multicolumn{7}{l}{\textbf{Baseline}} \\
Sens $\uparrow$ & 0.772 & 0.772 & 0.830 & 0.854 & 0.932 & 0.576 \\
Spec $\uparrow$ & 0.468 & 0.468 & 0.366 & 0.321 & 0.151 & 0.719 \\
PPV $\uparrow$  & 0.0029 & 0.0029 & 0.0026 & 0.0025 & 0.0022 & 0.0040 \\
Thr             & 0.00132 & 0.00132 & 0.00108 & 0.00095 & 0.00085 & 0.00217 \\
\bottomrule
\end{tabular}
\label{app:decision_thresholds}
\end{table}

\newpage

\section{Subgroup analysis}\label{app:subgroup_analysis}

\begin{table}[ht!]
\tiny
\setlength{\tabcolsep}{.8pt}
\caption{Subgroup performance of the Mix, Invasive, and Non-invasive models. Reported are AUROC, sensitivity, specificity, and event prevalence [pos/total], with thresholds targeting $\geq 80\%$ sensitivity on the test set.}
\begin{tabular}{lcccccccccccc} 
\toprule
& \multicolumn{4}{c}{\textbf{Mix}} & \multicolumn{4}{c}{\textbf{Invasive}} & \multicolumn{4}{c}{\textbf{Non-Invasive}} \\
\cmidrule(lr){2-5} \cmidrule(lr){6-9} \cmidrule(lr){10-13}
Feature & AUROC $\uparrow$ & Sens $\uparrow$ & Spec $\uparrow$ & [pos/total] & AUROC $\uparrow$ & Sens $\uparrow$ & Spec $\uparrow$ & [pos/total] & AUROC $\uparrow$ & Sens $\uparrow$ & Spec $\uparrow$ & [pos/total] \\
\midrule
gender: Male & 0.829 & 0.858 & 0.624 & [1349/660K] & 0.783 & 0.829 & 0.569 & [1051/316K] & \textbf{0.866} & 0.821 & 0.760 & [475/377K] \\
gender: Female & 0.809 & 0.842 & 0.595 & [863/456K] & 0.763 & 0.768 & 0.603 & [624/192K] & \textbf{0.830} & 0.746 & 0.775 & [339/291K] \\
ethnicity: White & 0.825 & 0.861 & 0.611 & [1579/831K] & 0.774 & 0.817 & 0.564 & [1194/379K] & \textbf{0.860} & 0.829 & 0.745 & [578/496K] \\
ethnicity: Black & 0.830 & 0.880 & 0.611 & [217/94.7K] & 0.757 & 0.830 & 0.529 & [165/40.3K] & \textbf{0.873} & 0.824 & 0.766 & [85/61.3K] \\
ethnicity: Asian & 0.784 & 0.765 & 0.689 & [34/35.5K] & \textbf{0.794} & 0.808 & 0.712 & [26/13K] & 0.711 & 0.500 & 0.816 & [10/23.7K] \\
ethnicity: Hispanic & 0.808 & 0.750 & 0.691 & [60/25.2K] & 0.788 & 0.698 & 0.668 & [43/10.2K] & \textbf{0.810} & 0.632 & 0.844 & [19/16.2K] \\
ethnicity: Other & 0.802 & 0.820 & 0.582 & [322/130K] & 0.782 & 0.785 & 0.608 & [247/66.5K] & \textbf{0.813} & 0.705 & 0.780 & [122/71.2K] \\
age $<$ 53.0 & 0.830 & 0.796 & 0.714 & [304/250K] & 0.797 & 0.747 & 0.683 & [245/104K] & \textbf{0.843} & 0.771 & 0.830 & [96/157K] \\
age [53.0 - 64.5) & 0.817 & 0.852 & 0.625 & [588/285K] & 0.771 & 0.758 & 0.655 & [451/138K] & \textbf{0.850} & 0.750 & 0.799 & [204/165K] \\
age [64.5 - 76.0) & 0.811 & 0.863 & 0.559 & [679/278K] & 0.758 & 0.793 & 0.577 & [492/133K] & \textbf{0.850} & 0.787 & 0.749 & [277/162K] \\
age $>$= 76.0 & 0.816 & 0.867 & 0.563 & [641/303K] & 0.777 & 0.836 & 0.528 & [487/134K] & \textbf{0.837} & 0.776 & 0.745 & [237/185K] \\
height $<$ 163.0 & 0.798 & 0.873 & 0.523 & [425/185K] & 0.751 & 0.834 & 0.458 & [326/86K] & \textbf{0.827} & 0.814 & 0.703 & [140/109K] \\
height [163.0 - 170.5) & 0.805 & 0.850 & 0.552 & [473/224K] & 0.782 & 0.799 & 0.603 & [373/114K] & \textbf{0.842} & 0.779 & 0.761 & [154/125K] \\
height [170.5 - 178.0) & 0.826 & 0.876 & 0.556 & [307/127K] & 0.774 & 0.801 & 0.589 & [251/64.7K] & \textbf{0.867} & 0.817 & 0.747 & [93/68.4K] \\
height $>$= 178.0 & 0.833 & 0.837 & 0.675 & [1007/581K] & 0.780 & 0.764 & 0.663 & [725/243K] & \textbf{0.859} & 0.754 & 0.811 & [427/366K] \\
weight $<$ 65.6 & 0.818 & 0.839 & 0.628 & [503/277K] & 0.769 & 0.822 & 0.524 & [349/111K] & \textbf{0.835} & 0.773 & 0.753 & [216/182K] \\
weight [65.6 - 79.6) & 0.828 & 0.869 & 0.602 & [542/285K] & 0.787 & 0.798 & 0.617 & [406/125K] & \textbf{0.877} & 0.817 & 0.766 & [186/173K] \\
weight [79.6 - 93.6) & \textbf{0.815} & 0.835 & 0.618 & [550/239K] & 0.776 & 0.773 & 0.637 & [449/119K] & 0.813 & 0.663 & 0.809 & [172/133K] \\
weight $>$= 93.6 & 0.826 & 0.864 & 0.602 & [617/316K] & 0.773 & 0.790 & 0.618 & [471/154K] & \textbf{0.877} & 0.825 & 0.779 & [240/181K] \\
bmi $<$ 23.4 & 0.805 & 0.863 & 0.566 & [365/173K] & 0.770 & 0.871 & 0.475 & [271/80.5K] & \textbf{0.847} & 0.803 & 0.735 & [132/104K] \\
bmi [23.4 - 27.7) & 0.810 & 0.877 & 0.557 & [473/189K] & 0.781 & 0.802 & 0.589 & [368/103K] & \textbf{0.834} & 0.758 & 0.766 & [157/97.4K] \\
bmi [27.7 - 32.1) & 0.801 & 0.845 & 0.553 & [470/188K] & 0.757 & 0.753 & 0.603 & [377/94.6K] & \textbf{0.829} & 0.752 & 0.740 & [153/105K] \\
bmi $>$= 32.1 & 0.834 & 0.838 & 0.664 & [904/566K] & 0.783 & 0.766 & 0.659 & [659/231K] & \textbf{0.864} & 0.772 & 0.806 & [372/363K] \\
obesity = 0 & 0.824 & 0.854 & 0.616 & [2024/1011K] & 0.778 & 0.825 & 0.560 & [1516/454K] & \textbf{0.851} & 0.807 & 0.747 & [767/613K] \\
obesity = 1 & 0.803 & 0.835 & 0.567 & [188/105K] & 0.751 & 0.780 & 0.591 & [159/54.3K] & \textbf{0.861} & 0.723 & 0.790 & [47/56K] \\
hypertension = 0 & 0.823 & 0.847 & 0.621 & [822/464K] & 0.778 & 0.807 & 0.574 & [615/213K] & \textbf{0.853} & 0.794 & 0.755 & [306/272K] \\
hypertension = 1 & 0.820 & 0.855 & 0.605 & [1390/653K] & 0.773 & 0.792 & 0.603 & [1060/295K] & \textbf{0.851} & 0.774 & 0.780 & [508/396K] \\
diabetes = 0 & 0.827 & 0.851 & 0.628 & [1459/776K] & 0.776 & 0.816 & 0.573 & [1092/345K] & \textbf{0.857} & 0.810 & 0.759 & [558/473K] \\
diabetes = 1 & 0.809 & 0.855 & 0.574 & [753/340K] & 0.772 & 0.806 & 0.582 & [583/163K] & \textbf{0.839} & 0.754 & 0.757 & [256/196K] \\
kidney disease = 0 & 0.826 & 0.851 & 0.629 & [1648/871K] & 0.782 & 0.818 & 0.581 & [1272/401K] & \textbf{0.850} & 0.785 & 0.770 & [585/519K] \\
kidney disease = 1 & 0.806 & 0.856 & 0.552 & [564/245K] & 0.746 & 0.794 & 0.533 & [403/107K] & \textbf{0.850} & 0.817 & 0.714 & [229/150K] \\
lung disease = 0 & 0.820 & 0.845 & 0.626 & [1442/772K] & 0.777 & 0.822 & 0.572 & [1083/346K] & \textbf{0.849} & 0.801 & 0.763 & [533/469K] \\
lung disease = 1 & 0.822 & 0.865 & 0.579 & [770/344K] & 0.771 & 0.789 & 0.590 & [592/162K] & \textbf{0.855} & 0.776 & 0.755 & [281/199K] \\
heart disease = 0 & 0.822 & 0.823 & 0.662 & [1295/727K] & 0.777 & 0.770 & 0.629 & [978/325K] & \textbf{0.853} & 0.776 & 0.785 & [482/442K] \\
heart disease = 1 & 0.817 & 0.893 & 0.519 & [917/389K] & 0.770 & 0.857 & 0.496 & [697/184K] & \textbf{0.844} & 0.813 & 0.720 & [332/227K] \\
drug abuse = 0 & 0.822 & 0.857 & 0.607 & [2133/1060K] & 0.776 & 0.831 & 0.554 & [1606/484K] & \textbf{0.850} & 0.807 & 0.743 & [802/635K] \\
drug abuse = 1 & 0.807 & 0.734 & 0.704 & [79/56.4K] & 0.769 & 0.623 & 0.699 & [69/24.4K] & \textbf{0.891} & 0.750 & 0.841 & [12/33.4K] \\
depression = 0 & 0.820 & 0.854 & 0.603 & [1904/942K] & 0.775 & 0.825 & 0.552 & [1453/439K] & \textbf{0.848} & 0.804 & 0.744 & [679/557K] \\
depression = 1 & 0.832 & 0.841 & 0.659 & [308/174K] & 0.780 & 0.743 & 0.658 & [222/69.7K] & \textbf{0.867} & 0.793 & 0.798 & [135/112K] \\
sedatives given = 0 & 0.814 & 0.742 & 0.736 & [997/812K] & 0.779 & 0.750 & 0.673 & [648/284K] & \textbf{0.843} & 0.743 & 0.812 & [470/564K] \\
sedatives given = 1 & 0.758 & 0.942 & 0.278 & [1215/304K] & 0.746 & 0.836 & 0.469 & [1027/224K] & \textbf{0.783} & 0.881 & 0.457 & [344/105K] \\
blood products  = 0 & 0.822 & 0.839 & 0.630 & [1888/1048K] & 0.775 & 0.806 & 0.584 & [1415/467K] & \textbf{0.851} & 0.796 & 0.759 & [707/637K] \\
blood products  = 1 & 0.753 & 0.929 & 0.326 & [324/68.5K] & 0.715 & 0.896 & 0.317 & [260/41.6K] & \textbf{0.806} & 0.860 & 0.530 & [107/31.7K] \\
antibiotics  = 0 & 0.831 & 0.851 & 0.634 & [1687/910K] & 0.789 & 0.832 & 0.569 & [1259/397K] & \textbf{0.852} & 0.791 & 0.766 & [631/560K] \\
antibiotics  = 1 & 0.779 & 0.857 & 0.516 & [525/206K] & 0.730 & 0.743 & 0.577 & [416/112K] & \textbf{0.841} & 0.842 & 0.697 & [183/108K] \\
anticoag/antiPLT = 0 & 0.825 & 0.854 & 0.619 & [1784/905K] & 0.780 & 0.832 & 0.560 & [1330/395K] & \textbf{0.850} & 0.803 & 0.752 & [675/560K] \\
anticoag/antiPLT = 1 & 0.808 & 0.843 & 0.581 & [428/211K] & 0.759 & 0.759 & 0.611 & [345/114K] & \textbf{0.858} & 0.806 & 0.762 & [139/109K] \\
neuromusc. block = 0 & 0.819 & 0.843 & 0.619 & [2081/1102K] & 0.773 & 0.813 & 0.569 & [1548/495K] & \textbf{0.849} & 0.801 & 0.748 & [790/667K] \\
neuromusc. block = 1 & 0.669 & 0.992 & 0.039 & [131/14.1K] & 0.696 & 0.945 & 0.167 & [127/13.5K] & \textbf{0.715} & 1.000 & 0.206 & [24/1.5K] \\
analgesics = 0 & 0.825 & 0.783 & 0.704 & [1203/822K] & 0.785 & 0.787 & 0.627 & [844/297K] & \textbf{0.846} & 0.757 & 0.798 & [490/561K] \\
analgesics = 1 & 0.771 & 0.935 & 0.354 & [1009/295K] & 0.756 & 0.826 & 0.519 & [831/212K] & \textbf{0.796} & 0.867 & 0.525 & [324/108K] \\
crystalloids = 0 & 0.903 & 0.733 & 0.916 & [101/137K] & 0.886 & 0.864 & 0.751 & [59/18.6K] & \textbf{0.914} & 0.675 & 0.940 & [40/120K] \\
crystalloids = 1 & 0.808 & 0.858 & 0.569 & [2111/979K] & 0.770 & 0.779 & 0.610 & [1616/490K] & \textbf{0.836} & 0.778 & 0.749 & [774/549K] \\
electrolytes = 0 & 0.830 & 0.834 & 0.655 & [1466/866K] & 0.777 & 0.799 & 0.594 & [1043/354K] & \textbf{0.859} & 0.796 & 0.764 & [603/555K] \\
electrolytes = 1 & 0.782 & 0.889 & 0.461 & [746/250K] & 0.765 & 0.843 & 0.532 & [632/155K] & \textbf{0.814} & 0.796 & 0.703 & [211/114K] \\
gi protection  = 0 & 0.825 & 0.855 & 0.615 & [2045/1037K] & 0.779 & 0.829 & 0.558 & [1537/470K] & \textbf{0.856} & 0.815 & 0.748 & [767/623K] \\
gi protection = 1 & 0.777 & 0.820 & 0.576 & [167/79.2K] & 0.737 & 0.732 & 0.612 & [138/38.1K] & \textbf{0.781} & 0.638 & 0.769 & [47/45.7K] \\
parenteral nutrition = 0 & 0.823 & 0.853 & 0.614 & [2054/1039K] & 0.774 & 0.829 & 0.551 & [1541/459K] & \textbf{0.850} & 0.805 & 0.747 & [778/635K] \\
parenteral nutrition = 1 & 0.802 & 0.842 & 0.585 & [158/77.2K] & 0.784 & 0.754 & 0.669 & [134/49.9K] & \textbf{0.887} & 0.806 & 0.786 & [36/33.5K] \\
antiarrhythmics = 0 & 0.823 & 0.842 & 0.626 & [2012/1069K] & 0.778 & 0.814 & 0.578 & [1510/476K] & \textbf{0.848} & 0.794 & 0.754 & [741/650K] \\
antiarrhythmics = 1 & 0.751 & 0.955 & 0.287 & [200/47.4K] & 0.712 & 0.903 & 0.322 & [165/32.1K] & \textbf{0.836} & 0.918 & 0.535 & [73/19.1K] \\
positive event = 0 & nan & nan & 0.612 & [0/1114K] & nan & nan & 0.558 & [0/507K] & nan & nan & 0.747 & [0/668K] \\
positive event = 1 & nan & 0.852 & nan & [2212/2.2K] & nan & 0.781 & nan & [1675/1.7K] & nan & 0.769 & nan & [814/814] \\
last $\leq 65$ & \textbf{0.780} & 0.967 & 0.207 & [1046/229K] & 0.690 & 0.999 & 0.019 & [694/81.5K] & 0.655 & 0.907 & 0.195 & [162/12.3K] \\
65 < last $\leq 70 $& \textbf{0.759} & 0.874 & 0.463 & [333/147K] & 0.679 & 0.861 & 0.304 & [274/67.2K] & 0.649 & 0.806 & 0.402 & [31/7.8K] \\
70 < last $\leq 100 $ & \textbf{0.803} & 0.707 & 0.758 & [737/650K] & 0.746 & 0.558 & 0.785 & [631/309K] & 0.676 & 0.634 & 0.629 & [71/34K] \\
last > 100  & 0.796 & 0.635 & 0.825 & [96/90.7K] & 0.779 & 0.447 & 0.906 & [76/50.8K] & \textbf{0.801} & 0.650 & 0.843 & [20/8.9K] \\
Comorbidities = 0 & 0.823 & 0.802 & 0.693 & [278/192K] & 0.782 & 0.757 & 0.658 & [206/77.5K] & \textbf{0.846} & 0.761 & 0.819 & [113/125K] \\
Comorbidities = 1 & 0.818 & 0.836 & 0.639 & [505/272K] & 0.773 & 0.764 & 0.642 & [365/125K] & \textbf{0.845} & 0.738 & 0.816 & [206/162K] \\
Comorbidities $\geq$ 2 & 0.820 & 0.868 & 0.577 & [1429/652K] & 0.773 & 0.801 & 0.582 & [1104/306K] & \textbf{0.854} & 0.792 & 0.752 & [495/382K] \\
\bottomrule
\end{tabular}
\label{app:subgroup_performance}
\end{table}

\newpage

\section{Shapley}\label{app:shapley}

\begin{figure}[ht!]  
\centering
\tiny
\includegraphics[width=0.8\textwidth, height=0.8\textheight, keepaspectratio]{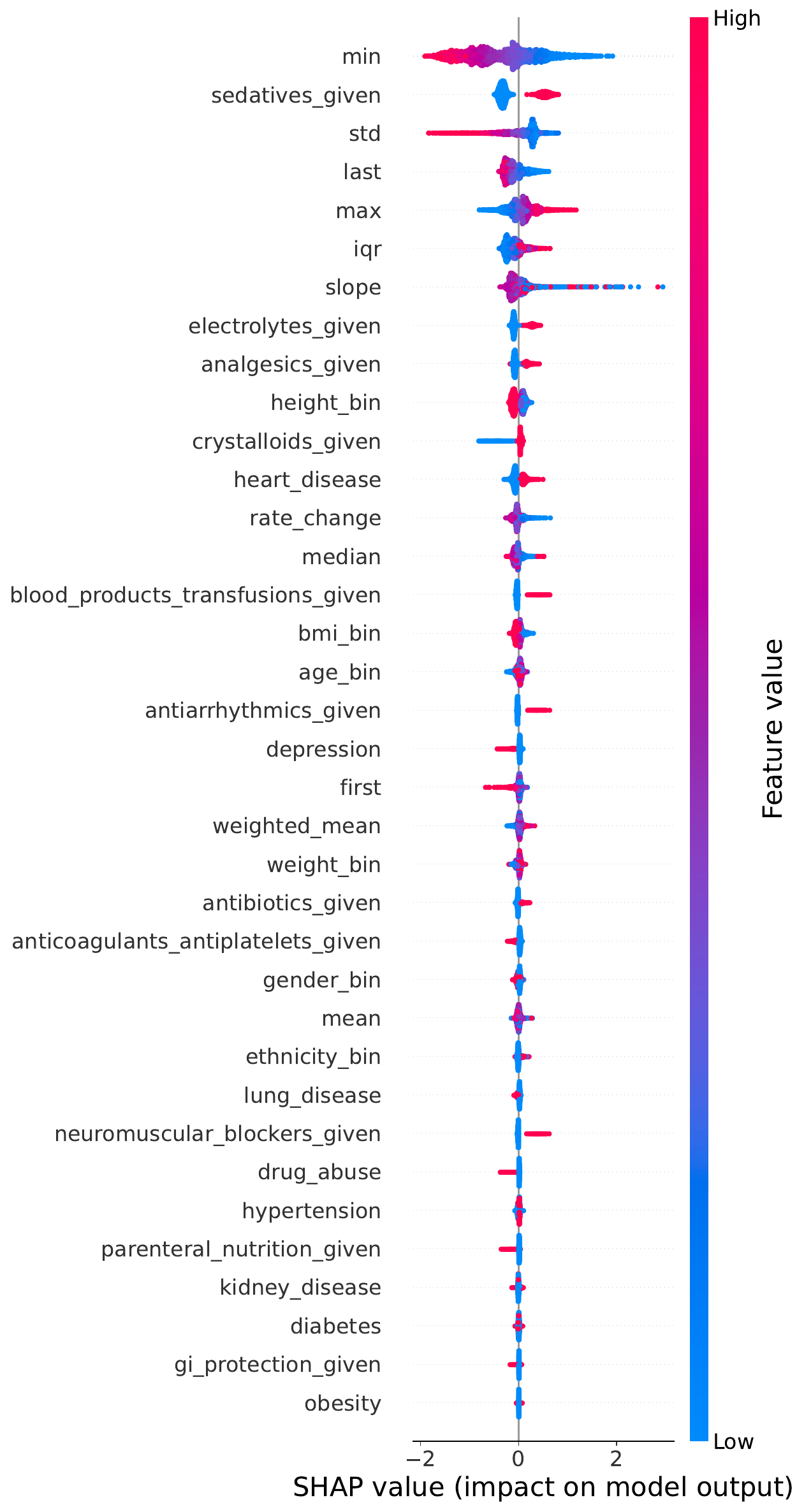}
\caption{SHAP feature importance plot for the \textbf{Mix} model.}
\label{app:shap_mix_full}
\end{figure}

\begin{figure}[ht!]
\centering
\tiny
\includegraphics[width=0.8\textwidth, height=0.8\textheight, keepaspectratio]{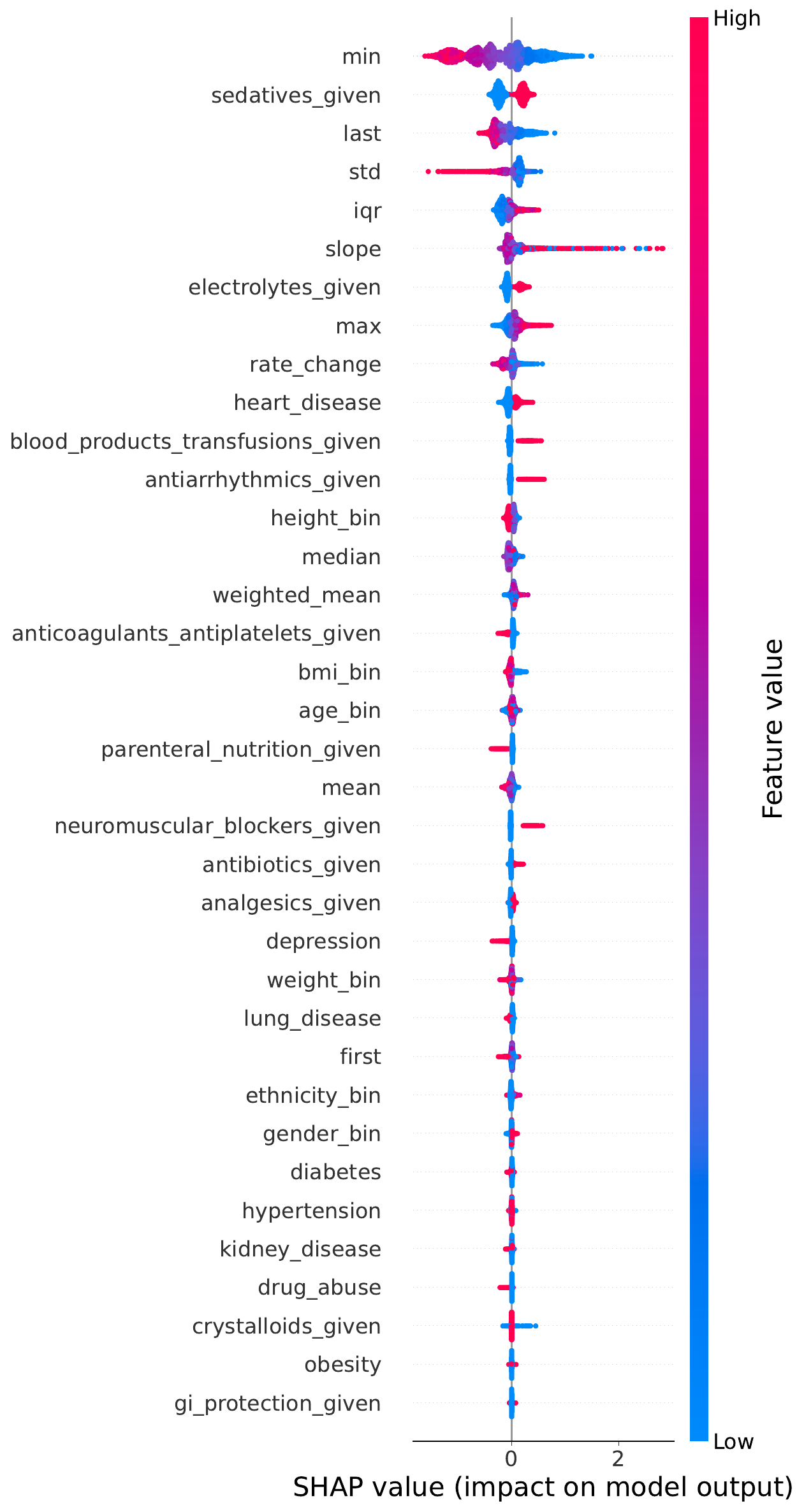}
\caption{SHAP feature importance plot for the \textbf{Invasive} model.}
\label{app:shap_inv_full}
\end{figure}

\begin{figure}[ht!]
\centering
\tiny
\includegraphics[width=0.8\textwidth, height=0.8\textheight, keepaspectratio]{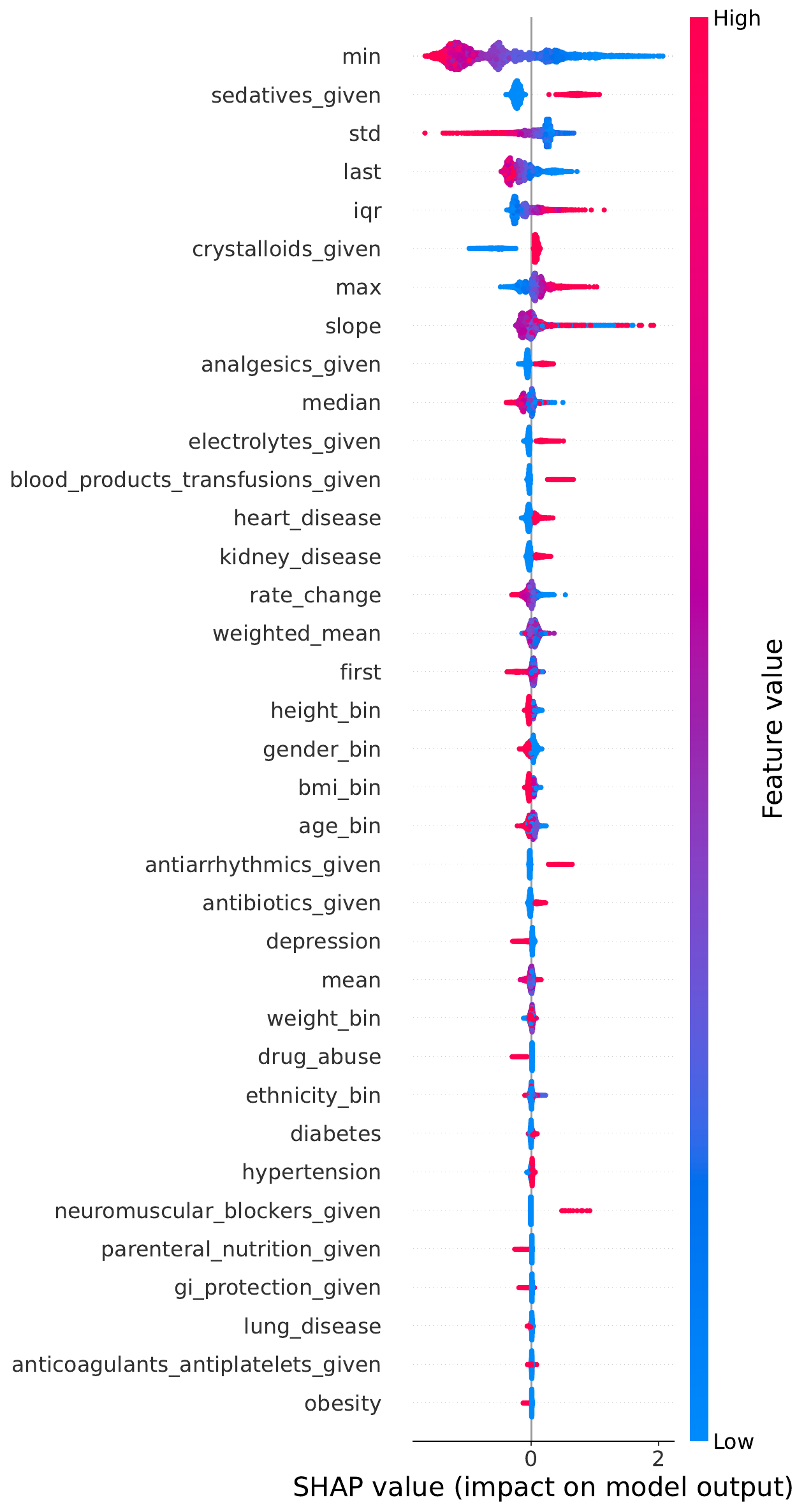}
\caption{SHAP feature importance plot for the \textbf{Non-Invasive} model.}
\label{app:shap_noninv_full}
\end{figure}

\end{appendices}

\clearpage

\end{document}